# Spectroscopic Studies of the Physical Origin of Environmental Aging Effects on Doped Graphene

J.-K. Chang,[1,2] C.-C. Hsu,[1] S.-Y. Liu,[2] C.-I Wu,[2] M. Gharib,[3] and N.-C. Yeh[1,4]*

[1]*Department of Physics, California Institute of Technology, Pasadena, CA 91125, USA*

[2]*Graduate Institute of Photonics and Optoelectronics, Department of Electrical Engineering, National Taiwan University, Taipei, Taiwan 10617*

[3]*Department of Aeronautics, California Institute of Technology, Pasadena, CA 91125, USA*

[4]*Kavli Nanoscience Institute, California Institute of Technology, Pasadena, CA 91125, USA*

(*Author to whom correspondence should be addressed. Electronic mail: ncyeh@caltech.edu)

The environmental aging effect of doped graphene is investigated as a function of the organic doping species, humidity, and the number of graphene layers adjacent to the dopant by studies of the Raman spectroscopy, x-ray and ultraviolet photoelectron spectroscopy, scanning electron microscopy, infrared spectroscopy, and electrical transport measurements. It is found that higher humidity and structural defects induce faster degradation in doped graphene. Detailed analysis of the spectroscopic data suggest that the physical origin of the aging effect is associated with the continuing reaction of $H_2O$ molecules with the hygroscopic organic dopants, which leads to formation of excess chemical bonds, reduction in the doped graphene carrier density, and proliferation of damages from the graphene grain boundaries. These environmental aging effects are further shown to be significantly mitigated by added graphene layers.

**I. INTRODUCTION**

Among many technological promises of graphene,[1-3] the feasibility of employing graphene as transparent conducting electrodes in optoelectronic devices has been demonstrated.[4-6] In particular, the application of graphene-based electrodes to organic photovoltaic cells (OPVCs) and organic light emitting diodes (OLEDs) are especially appealing because of the potential of achieving inexpensive and flexible optoelectronic devices.[6] In the case of graphene-based OPVCs, reasonable power conversion efficiency has been realized.[7,8] However, apparent degrading performance with time has been a major challenge for practical applications.[6] It has been reported that stacked multi-layers of graphene with a spacing less than 0.7 nm atop organic molecules could protect the OPVC from rapid decrease in the power conversion efficiency within the first three days,[6] and that coating of multi-layers of graphene on Cu or Ni could prevent the metal from rapid oxidation.[9] Moreover, stacked graphene oxide layers were found to be water impermeable.[10] Nonetheless, the underlying causes and reliable remedies for the aging effects of doped graphene have not been systematically investigated.

Here we report systematic imaging and spectroscopic studies of both hole and electron doped graphene as a function of time for up to 30 days under different conditions, and compare these results with those of undoped, pristine graphene. We find that the aging effect is more severe if the ambient humidity is higher, and that faster aging effects occur along the defects in graphene. Moreover, organic molecules intercalated between two sheets of monolayer graphene appear to be much more stable than those covered under a monolayer graphene sheet. Our experimental findings can be consistently explained by attributing the environmental aging effect to continuing reaction of $H_2O$ molecules with the hygroscopic organic dopants, which lead to propagating damages along the graphene grain boundaries and continuous decrease in the graphene carrier densities due to the degradation of organic dopants.

## II. EXPERIMENTAL

We conducted time-dependent spectroscopic and imaging studies of doped graphene samples prepared under six different conditions, as summarized in Fig. 1(a). In addition, controlled pristine graphene samples without organic doping were studied as a function of time for comparison. The monolayer graphene films used for our investigation were $(8\times8)$ mm$^2$ in size and were synthesized by the chemical vapor deposition (CVD) process on copper foils, all purchased from Graphene Supermarket. The graphene films were transferred from their original copper substrates by a polymer-free technique,[11] as schematically shown in Fig. 1(b).

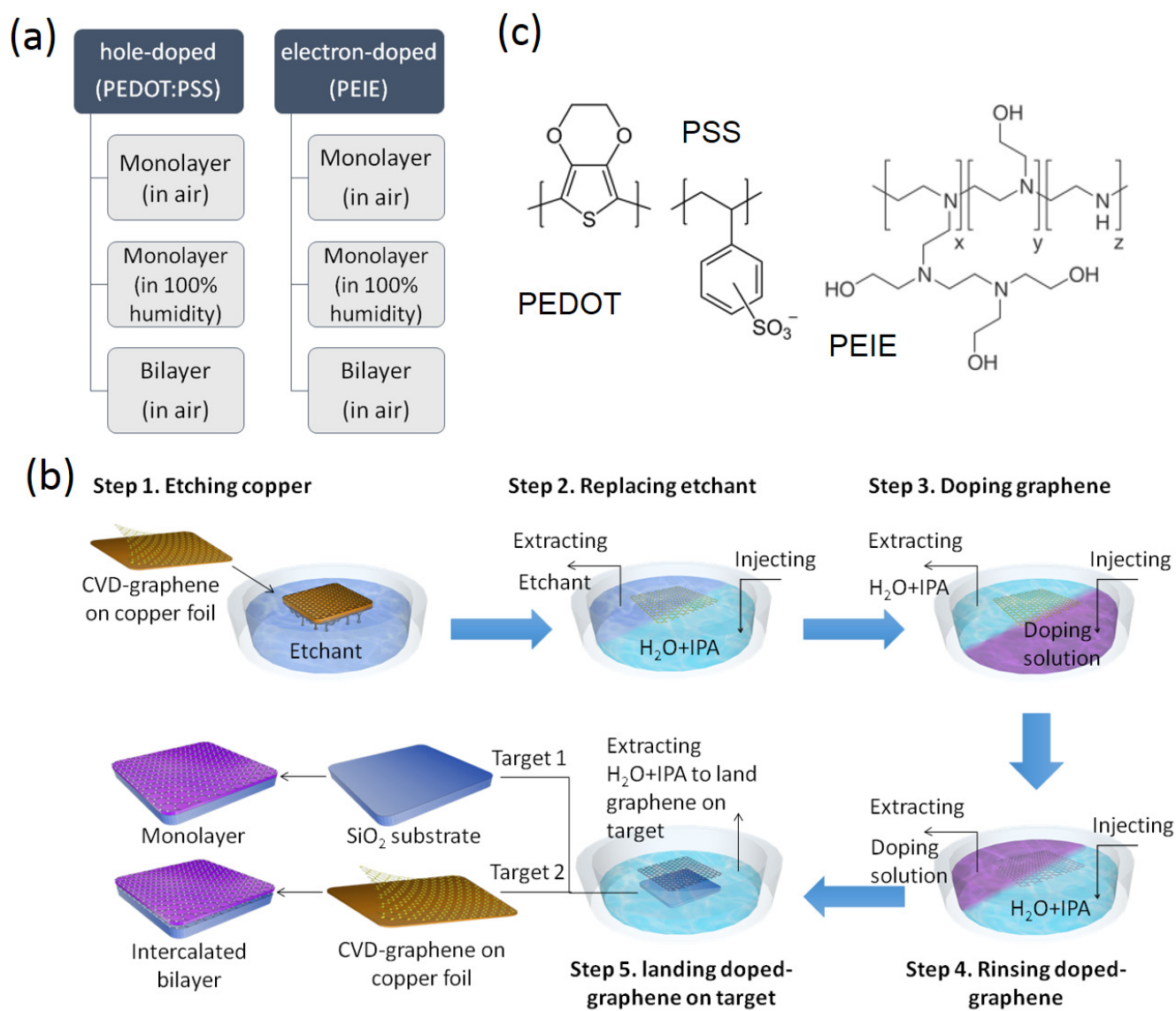

FIG. 1. Conditions and preparation procedures of doped graphene samples used in this work: (a) Doped graphene samples prepared under six different conditions were investigated as a function of time. (b) Schematics of preparation procedures of the doped graphene samples, where doped monolayer graphene is prepared by following the "Target 1" process, whereas intercalated bilayer graphene is prepared by following the "Target 2" process. (c) The molecular structures for the hole-dopant PEDOT and the electron-dopant PEIE.

To prepare doped monolayer graphene, a graphene-on-copper sample was first placed in the middle of a graphite confinement and immersed in the ammonium persulfate etchant (Alfa Aesar, 0.2M) to etch away the copper substrate. After the copper foil was removed, the pristine graphene film was buoyed up by an aqueous solution (a mixture of deionized water and IPA in the ratio of 10:1) that replaced the etching waste. Meanwhile, a dilute PEDOT:PSS solution (Heraeus, CleviosTM PH 1000) for hole-doping, or a dilute PEIE solution (Alderich, 0.05wt%) for electron-doping, was gradually introduced into the aqueous solution to obtain a doped graphene by liquid phase diffusion beneath the graphene sheet. (See Fig. 1(c) for the molecular structures of PEDOT:PSS and PEIE.) Then the doped solution was completely rinsed off by the IPA-deionized water mixture prior to placing a substrate under the doped graphene to ensure that the substrate was free of residue. Finally, the IPA-deionized water mixture was pumped out to land the doped graphene onto the "target" (typically a $SiO_2$/Si substrate), and the doped graphene was subsequently heated at 60 °C for 10 minutes to improve the adhesion and to eliminate residue moisture.

To obtain the intercalated bilayer with the dopant incorporated between two graphene sheets, the aforementioned polymer-free transfer and doping process was first carried out with a monolayer graphene-on-copper as the target, which led to the graphene-dopant-graphene-copper assembly. The copper was then removed and another substrate (e.g. $SiO_2$/Si) was introduced so that the final sample became intercalated bilayer graphene on $SiO_2$/Si, as shown in Fig. 1(b).

Raman spectroscopy, x-ray photoelectron spectroscopy (XPS) and ultraviolet photoelectron spectroscopy (UPS) were performed on all graphene samples. The Raman spectra were taken with a Renishaw M1000 micro-Raman spectrometer system using a 514.3 nm laser (2.41 eV) as the excitation source. A 50× objective lens with a numerical aperture of 0.75 and a 2400 lines/mm grating were chosen during the measurement to achieve better signal-to-noise ratio. For each sample after a given aging time, the Raman spectra were taken at nine different locations. The resulting spectra were analyzed, and all spectral characteristics were obtained by using the average values as the mean and the range of variations as the empirical error. XPS was performed under $10^{-9}$ Torr with a Surface Science Instruments M-Probe that utilized Al $K_\alpha$ X-rays and a hemispherical energy analyzer. The UPS experiments were carried out in a Phi5400 system with a base pressure of $10^{-10}$ Torr. During the UPS measurement with He II (photon energy $h\nu$ = 40.8 eV) as the excitation source, the photoelectrons were collected by a hemispherical analyzer with an energy resolution of 0.05 eV. Time-evolved scanning electron microscopic (SEM) images were also taken on all graphene samples by a FEI Nova 600 SEM system with the following parameters: acceleration voltage = 5 kV, beam current = 98 pA, and working distance ~ 5 mm. Additionally, Fourier transformed infrared (FT-IR) spectroscopic studies were carried out on PEIE using an FT-IR spectrometer (Nexus 470). Both transmittance and absorbance spectra were obtained over the spectral range from 400 cm$^{-1}$ to 4000 cm$^{-1}$.

For studies of the aging effect in air, we verified that the averaged humidity in Pasadena during the period of our experiments varied from 21% to 34% with a mean value of 28% according to the official data from the National Oceanic and Atmospheric Administration. All "in-air" graphene samples were stored in an air-condition laboratory where the humidity was lower and varying less than the external humidity, ranging from 18% to 28% with a mean value of 23% over the same period of time. These samples were only exposed to external air briefly whenever they were transported to another building on campus for XPS and UPS measurements. Therefore, the humidity values for the "in-air" condition may estimated at ~ (23±5)%. To investigate doped graphene with saturated humidity, a CSZ MicroClimate environmental testing chamber was used for sample storage to simulate a saturated humidity condition, and the accuracy of controlled humidity was found to be ~ (95±5)% for ambient temperatures maintained at (25±3)°C.

## III. RESULTS & ANALYSIS

The electrical transport properties of the as-doped graphene and pristine graphene samples were characterized using the field-effect-transistor (FET) configuration at room temperature, confirming excess holes (~ $3.24 \times 10^{12}$ cm$^{-2}$) in the PEDOT:PSS-doped graphene and excess electrons (~ $2.31 \times 10^{12}$ cm$^{-2}$) in the PEIE-doped graphene, as shown by the gate voltage shifts in Fig. 2(a)-(b) relative to the Dirac point of pristine graphene and further corroborated by the UPS studies of the work functions shown in Fig. 2(c). Additionally, comparison of the UPS data taken on the as-doped graphene samples and on the same samples after one-month in air is shown in Fig. 2(d), revealing decrease (increase) in the work function (defined as the photon energy minus the binding energy) by 0.15 eV (0.2 eV) in the hole- (electron-) doped graphene, which corresponded to decreasing hole (electron) carriers with time. The decreasing hole-doping level due to the reaction of moisture with the hole dopant may be understood by the known process[12-14]: $H_2O + PSS(HSO_3^+) \rightarrow H_3O^+ + PSS(SO_3)$ so that the hole doping from PSS to graphene becomes reduced by the presence of water molecules. Similarly, the decreasing electron-doping level due to the reaction with moisture may be understood by the process $H_2O + PEIE\ (OH^-) \rightarrow OH^- + PEIE\ (H_2O)$ so that the electron doping from PEIE to graphene is reduced.

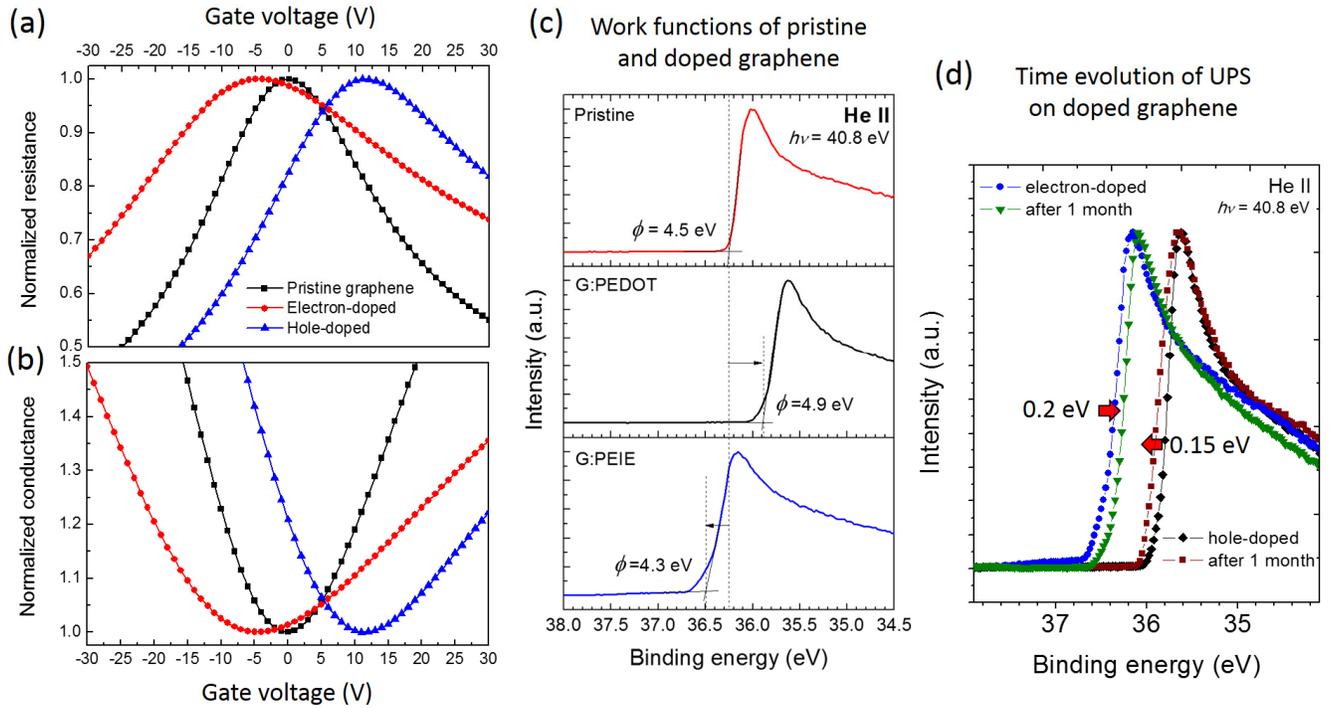

FIG. 2. Graphene Fermi level shifts with dopant and time: (a) FET measurements of the normalized resistance as a function of the gate voltage (V) relative to the Dirac point of the pristine graphene, where the thickness of the SiO$_2$ gate dielectric was 300 nm and the voltage shift of the maximum resistance corresponded to a carrier density of $3.24 \times 10^{12}$ cm$^{-2}$ ($2.31 \times 10^{12}$ cm$^{-2}$) for the hole-doped (electron-doped) graphene, and the maximum sheet resistance values for the pristine, electron-doped and hole-doped graphene samples were 540.2 Ω/□, 283.4 Ω/□ and 657.1 Ω/□, respectively; (b) FET measurements of the normalized conductance as a function of the gate voltage (V) relative to the Dirac point of the pristine graphene, where the minimum sheet conductance values for the pristine, electron-doped and hole-doped graphene samples were $1.82 \times 10^{-3}$ S, $3.54 \times 10^{-3}$ S and $1.51 \times 10^{-3}$ S, respectively; (c) UPS studies of the work functions ($\phi$) of pristine and organically doped graphene, with He II as the excitation source: Top panel, pristine graphene showing $\phi$ = 4.5 eV; Middle panel, graphene freshly doped with PEDOT:PSS showing $\phi$ = 4.9 eV, consistent with hole doping; Bottom panel: graphene freshly doped with PEIE showing $\phi$ = 4.3 eV, consistent with electron doping. (d) Time dependent UPS of the hole- (electron-) doped graphene stored in air, showing that the work function decreased (increased) from 5.1 eV (4.2 eV) to 4.95 eV (4.4 eV) after one month, consistent with reduced hole- (electron-) carriers with increasing time.

The time evolution of the Raman spectral characteristics for the 2D-, G- and D-bands[15] of all doped graphene samples under 6 different conditions is shown in Fig. 3(a)-(f) and the initial values of the spectral characteristics of the as-doped samples are summarized in Table I. Additionally, the spectral characteristics of as-grown graphene are included in Table I for comparison.

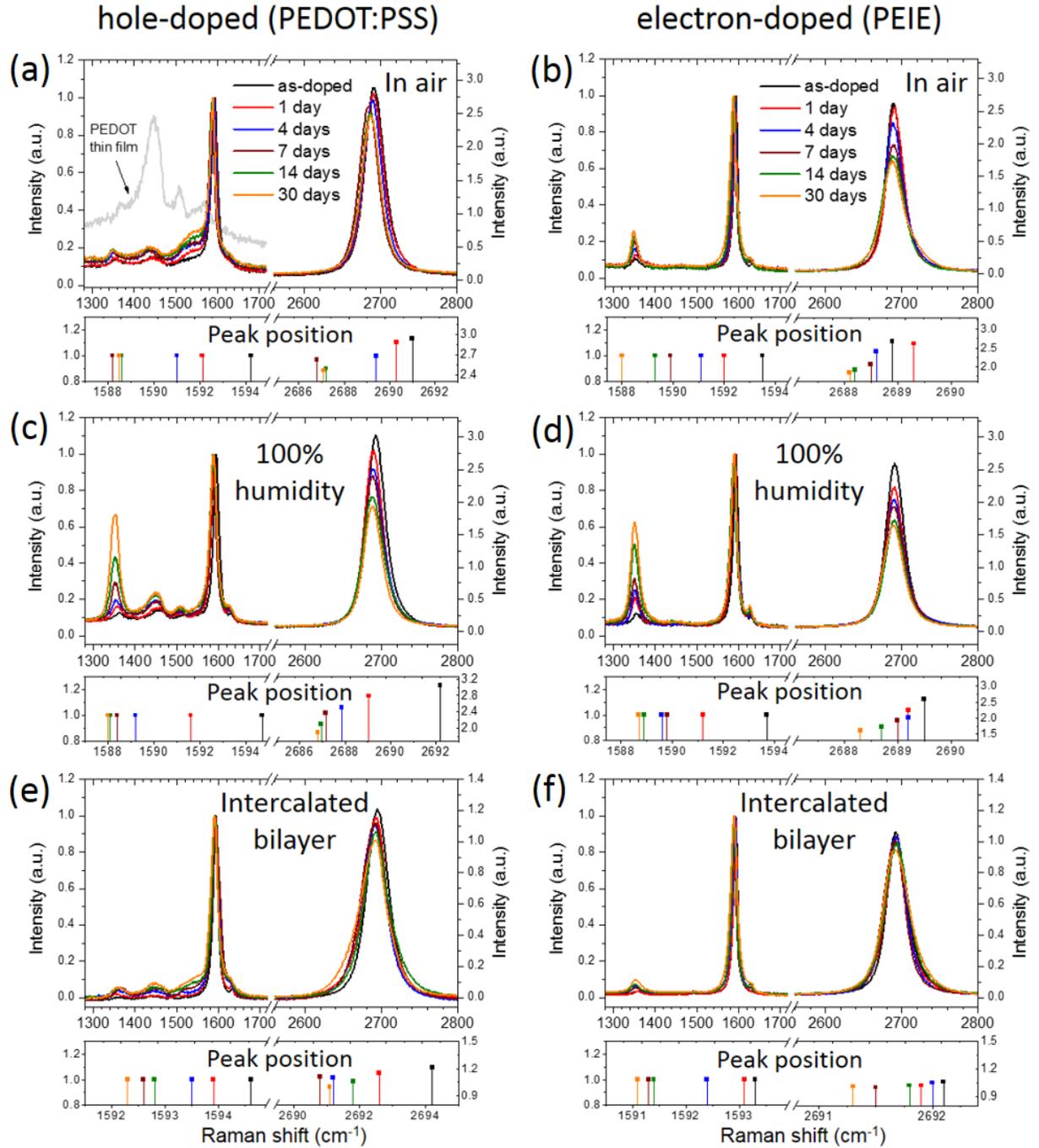

FIG. 3. The time-dependent Raman spectral evolution of doped graphene samples, with the peak positions of the graphene Raman modes specified in the lower panels for easy comparison: (a) PEDOT:PSS hole-doped monolayer graphene in air; (b) PEIE electron-doped monolayer graphene in air; (c) PEDOT:PSS hole-doped monolayer graphene in 100% humidity; (d) PEIE electron-doped monolayer graphene in 100% humidity; (e) PEDOT:PSS-intercalated bilayer graphene in air; (f) PEIE-intercalated bilayer graphene in air. Here the excess spectral feature near ~ 1450 cm$^{-1}$ that only appeared in the hole-doped graphene and increased with time is consistent with the Raman mode of the PEDOT:PSS dopant,[18,19] as shown in (a).

Table I. Raman spectral characteristics of as-doped monolayer and intercalated bilayer graphene

|  | PEDOT:PSS (hole)-doped monolayer | PEDOT:PSS (hole)-intercalated bilayer | PEIE (electron)-doped monolayer | PEIE (electron)-doped bilayer | Pristine graphene |
|---|---|---|---|---|---|
| **2D-band peak position (cm$^{-1}$)** | (2691.7±1.6) | (2694.0±1.2) | (2689.1±1.1) | (2692.4±1.4) | (2690.3±2.7) |
| **G-band peak position (cm$^{-1}$)** | (1594.4±2.5) | (1594.7±2.0) | (1593.5±1.2) | (1593.3±1.0) | (1587.6±2.4) |
| **FWHM of 2D-band (cm$^{-1}$)** | (29.0±2.5) | (34.1±1.5) | (26.5±2.4) | (34.2±1.8) | (25.6±2.9) |
| **FWHM of G-band (cm$^{-1}$)** | (16.8±0.9) | (17.1±1.1) | (10.9±1.0) | (11.7±1.1) | (11.2±1.0) |
| ($I_{2D}/I_G$) | (2.89±0.20) | (1.19±0.11) | (2.48±0.14) | (1.06±0.11) | (3.90±0.10) |
| ($I_D/I_G$) | (0.15±0.03) | (0.10±0.03) | (0.12±0.02) | (0.08±0.04) | (0.12±0.03) |

In general, all doped graphene samples exhibit the following common time evolution regardless of the conditions. First, the 2D-band that represents a double-resonance process of a perfect monolayer graphene exhibited steadily decreasing peak intensity and increasing linewidth with time. Second, the peak position of the G-band, which is associated with the doubly degenerate zone-center $E_{2g}$ mode of graphene, downshifted with time while the intensity remained the same. Third, the intensity of the D-band, which is associated with the zone-boundary phonons due to defects, increased strongly with time. Fourth, the full-width-half-maximum (FWHM) linewidth of all three Raman modes increased steadily with time.

To better understand how varying conditions influenced the aging of graphene, we normalize the following time-dependent spectral characteristics to their initial "as-doped" values: the peak positions of the 2D-and G-bands, the FWHM linewidths of the 2D- and G-bands, and the intensity ratios ($I_{2D}/I_G$) and ($I_D/I_G$) of 2D-to-G and D-to-G, respectively. Our rationale for choosing these quantities is that increasing FWHM linewidths of the 2D- and G-bands, decreasing values of ($I_{2D}/I_G$), and increasing values of ($I_D/I_G$) are generally indicative of the degradation of monolayer graphene quality.[16,17] Specifically, the magnitude of ($I_D/I_G$) represents the deviation from the perfect $sp^2$ hybridization of monolayer graphene as the result of defects/edges or the formation of chemical bonds between graphene π-electrons and other molecules. In fact, the intensifying ($I_D/I_G$) ratio in Figs. 3(a)-(d) with time for doped monolayer graphene may be attributed to increasing C-OH, C-O or C-C=O bonds and also increasing edge states, as corroborated by our XPS studies and SEM images to be discussed later. Similarly, the decreasing ($I_{2D}/I_G$) ratio coupled with increasing FWHM linewidth of the 2D-band imply broadening of the phonon modes involved in the double resonance of monolayer graphene. The broadening can be understood as the result of symmetry-breaking among the carbon atoms due to excess edge states and/or formation of other chemical bonds that are mediated by the reaction of organic dopants with moisture, leading to broadening of the phonon energies involved in this fourth-order process.

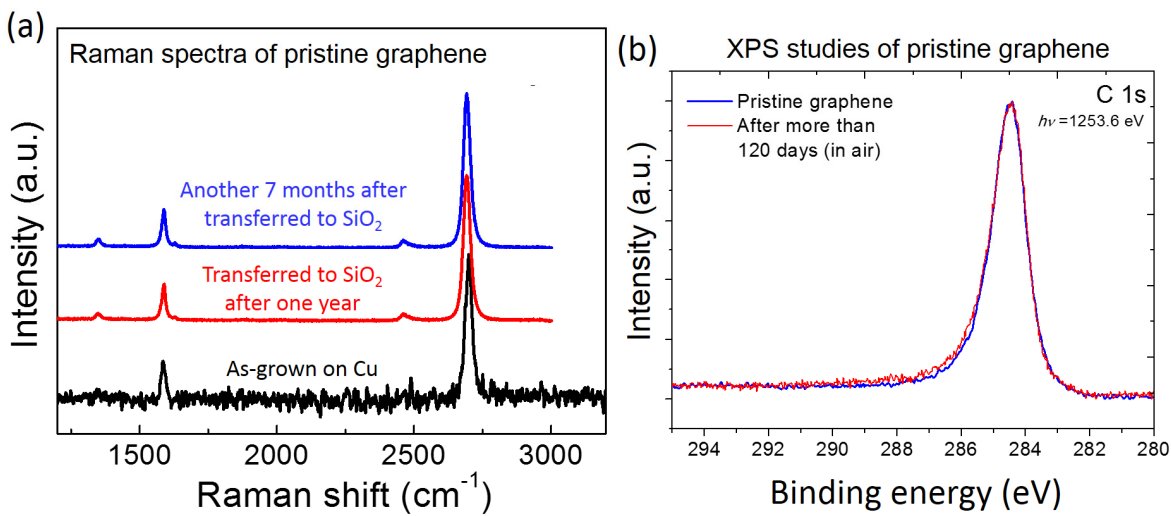

FIG. 4. Comparison of the Raman spectra and XPS of an undoped piece of PECVD-grown graphene taken at different times after growth: (a) From bottom to top, Raman spectra of graphene taken on Cu substrate immediately after CVD-growth (black), on the same graphene sample that was transferred to SiO$_2$/Si substrate after one year of its growth (red), and on the same graphene sample after another 7 months on the SiO$_2$/Si substrate (blue). (b) XPS studies of pristine graphene that was grown on Cu and transferred to SiO$_2$, showing no discernible changes after stored in air for four months.

For comparison, no discernible aging effects can be found in either the Raman spectra or XPS studies of undoped CVD-grown graphene stored in air, as exemplified in Fig. 4. These findings suggest that the reaction between the organic dopant and environmental moisture is the primary cause for the aging effect in doped graphene.

In Fig. 5(a)-(l) we investigate how varying humidity (first and third columns) and the number of graphene layers (second and fourth columns) influence the aging effect of both hole-doped and electron-doped monolayer graphene.

Concerning the effect of varying humidity, we find that for both monolayer graphene samples doped with holes and electrons, the peak positions of the 2D- and G-bands (left panels of Fig. 5(a)-(b) and 5(g)-(h)) exhibited initial downshift with time and then became saturated within experimental errors. In particular, the initial downshift of the G-band frequency appeared faster in 100% humidity. Similarly, the FWHM of both the 2D- and G-bands (left panels of Fig. 5(c)-(d) and 5(i)-(j)) exhibited more rapid initial increase in 100% humidity followed by an eventual saturation. As we shall substantiate with additional information later, the initial downshifts in the frequencies of the 2D- and G-bands may be attributed to the increasing phonon masses due to moisture-induced formation of extra bonds to the carbon atoms.

In contrast, the D-to-G intensity ratios ($I_D/I_G$) of both the hole- and electron-doped graphene exhibited continuous increase without saturation in 100% humidity and much smaller variations with time if stored in air (left panels of Fig. 5(f) and 5(l)), whereas the 2D-to-G intensity ratios ($I_{2D}/I_G$) initially decreased rapidly but eventually saturated with time (left panels of Fig. 5(e) and 5(k)). We attribute the continuously increasing ($I_D/I_G$) ratios of doped monolayer graphene in 100% humidity to the continuously increasing edge states (from the SEM micrographs given later in this section) and the increasing formation of C-OH, C-O or C-C=O bonds (from XPS studies shown later in this section) due to the hygroscopic nature of both PEDOT:PSS and PEIE dopants that steadily attracted and reacted with the water molecules that went through the grain boundaries and defects of monolayer graphene. On the other hand, the ($I_{2D}/I_G$) ratio is associated with a fourth-order, double-resonance process and so is less sensitive to the humidity level than the first-order process associated with the D-band.

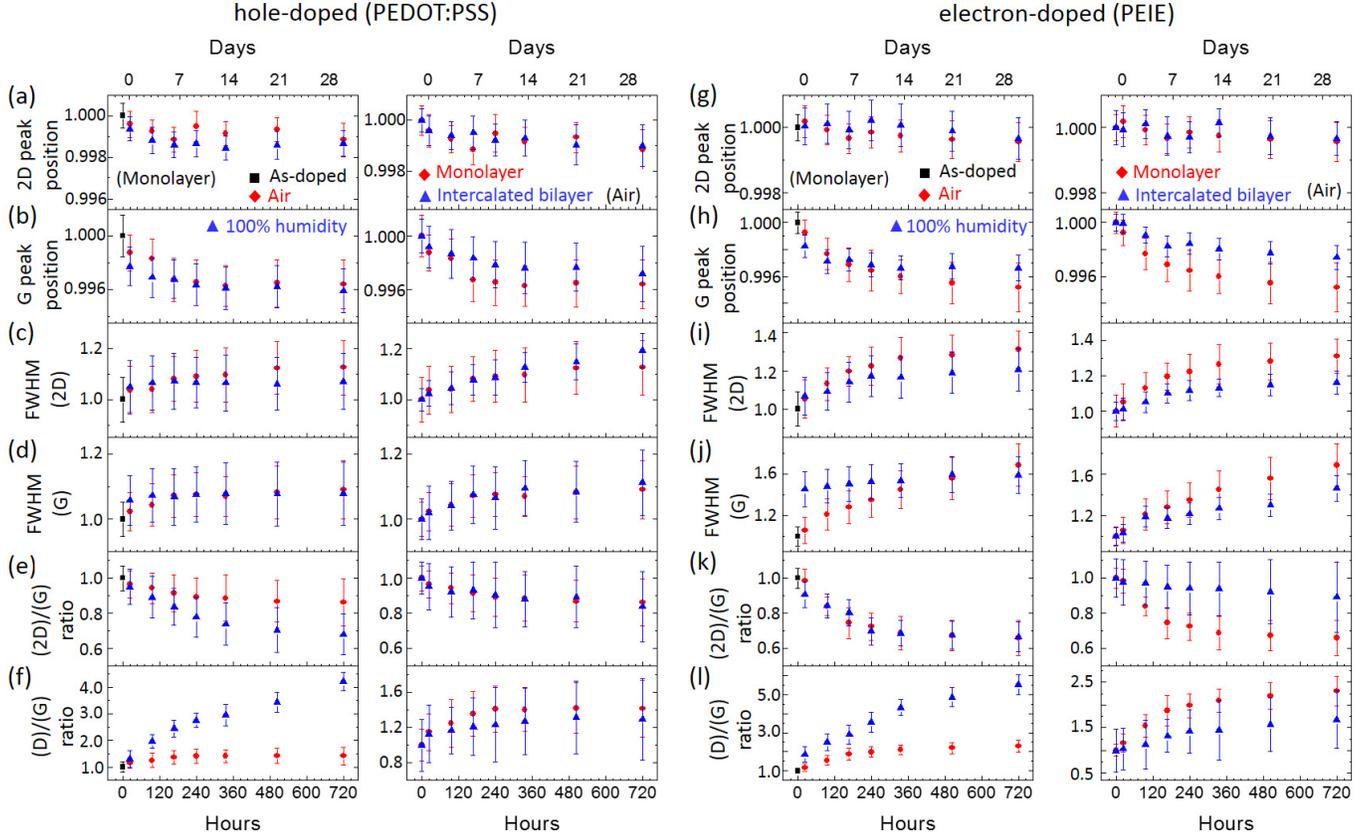

FIG. 5. Comparison of the effect of varying humidity (first and third columns) and varying number of graphene layers (second and fourth columns) on the aging of hole- and electron-doped graphene, from as-doped to 30 days: (a) & (g) normalized 2D-band peak position; (b) & (h) normalized G-band peak position; (c) & (i) normalized FWHM of the 2D-band; (d) & (j) normalized FWHM of the G-band; (e) & (k) normalized intensity ratio ($I_{2D}/I_G$); (f) & (l) normalized intensity ratio ($I_D/I_G$). In general the most pronounced aging effect of doped monolayer graphene due to humidity is manifested by the increasing ($I_D/I_G$) intensity ratio, and the aging effect becomes mitigated in the intercalated bilayer graphene.

Next, we consider the dependence of the aging effect of both hole- and electron-doped graphene on the number of graphene layers (right panels of Fig. 5(a)-(l)). We found that all spectral characteristics exhibited much less time-dependent variations and therefore better stability for the intercalated bilayer graphene (blue symbols) than the doped monolayer graphene (red symbols), and the benefits of the intercalated bilayer configuration were more significant in the electron-doped samples.

To better illustrate the differences of the aging effect between hole- and electron-doped graphene, we compare in Fig. 6(a)-(f) the time-dependent Raman spectral characteristics of hole- and electron-doped graphene stored in air (left panels), in 100% humidity (middle panels) and for intercalated bilayers in air (right panels). For monolayer graphene stored in either air or 100% humidity, most of the spectral characteristics for the hole-doped samples, except the peak positions of the 2D- and G-bands, exhibited better stability with time than the electron-doped samples. While both organic dopants are known to be hygroscopic based on studies of PEDOT:PSS in the literature[12-14] and our own FT-IR (Fourier transformed infrared) spectroscopic studies of the PEIE shown in Figure 7, the more significant time-dependent spectral degradation in electron-doped graphene suggested that the reaction of PEIE to moisture was more significant than that of PEDOT:PSS, which may be the result of a large number of $(OH)^-$ ions in PEIE (Fig. 1(c)) that continuously reacted to environmental moisture.

We further investigated the XPS of all doped graphene samples aged under various conditions. As shown in Fig. 8(a) for hole-doped monolayer graphene in air, the binding energy of the C-1s core electron appeared to increase with time up to 0.15 eV after 30 days, which was accompanied by increasing strengths of the $sp^2$ hybridized C-OH (at 285.7 eV) and the $sp^3$ C-O (at 286.6 eV) components of a comparable binding energy[21,22] as indicated by the arrow. The formation of the C-OH and C-O bonds may be understood as the result of the reaction of $H_3O^+$ with graphene, leading to either $H_3O^+ + C \rightarrow 2H^+ + (C-OH)^-$ or $H_3O^+ + C \rightarrow 3H^+ + (C-O)^{2-}$. Additionally, the binding energy of the S-2p core electron also increased with time up to 0.4 eV after 30 days, which may be attributed to the degradation of PEDOT:PSS because of the reaction of PSS with water that we have described earlier. A similar situation due to reactions of the water-soluble PEIE with moisture in air was also found in the case of electron-doped graphene shown in Fig. 8(b), where the spectrum of the binding energy of C-1s core electron exhibited an increase of the C-C=O component (at 288.7 eV)[21,22] with time plus a small trace of the C-OH component (at 285.7 eV). The relevant chemical reaction here may involve either $OH^- + C \rightarrow (C-OH)^-$ or $2OH^- +$ $2C \rightarrow H_2O + (C-C=O)^{2-}$. Further, both the binding energies of the C-1s and N-1s core electrons were downshifted by 0.2 eV in air after 30 days, suggesting that the work function increased by 0.2 eV, in excellent agreement with the UPS studies shown in Fig. 2.

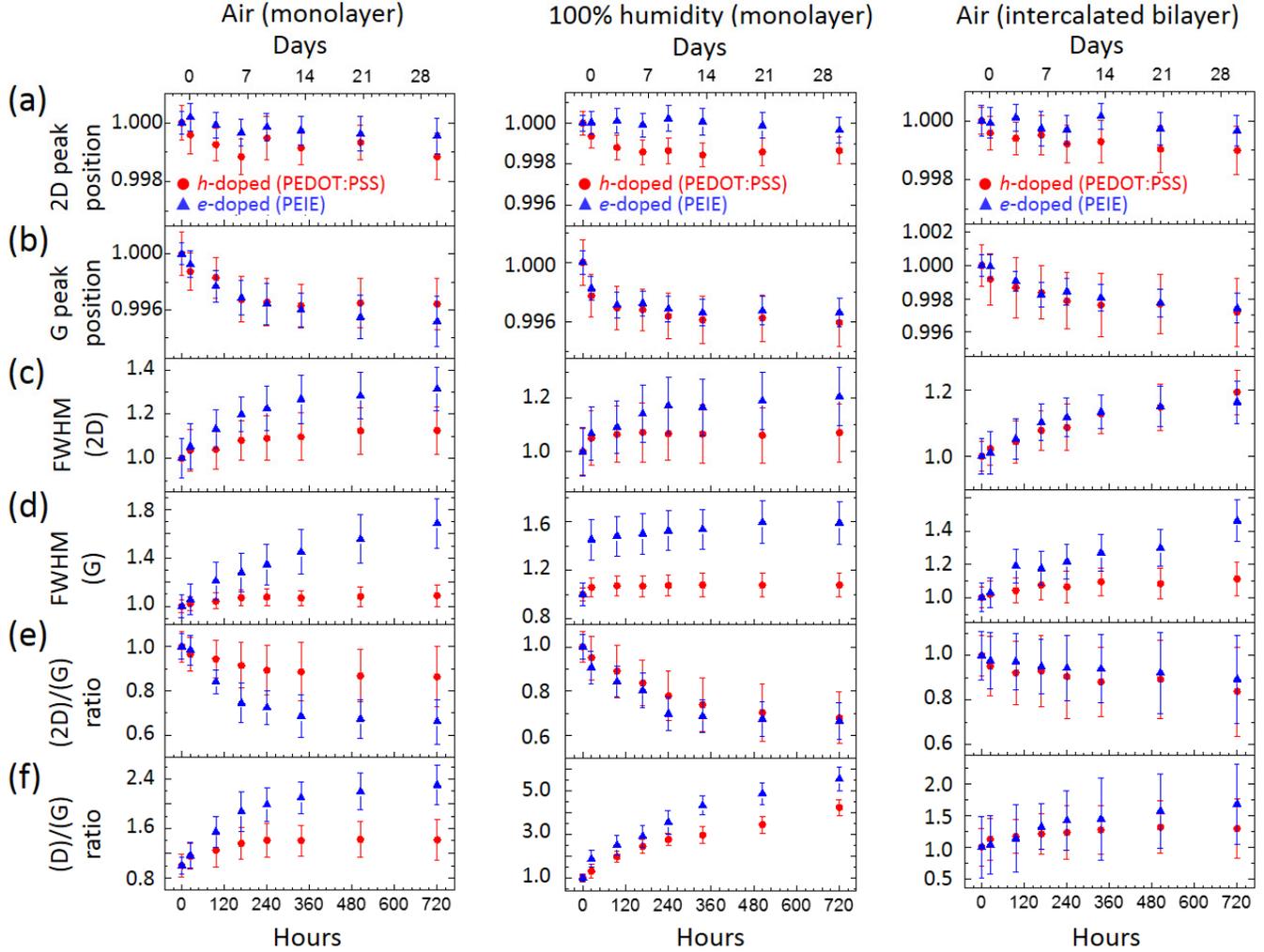

FIG. 6. Comparison of the effect of hole- vs. electron-doping on the aging of graphene in air (left panels), in 100% humidity (middle panels) and for intercalated bilayers in air (right panels) from as-doped to 30 days: (a) normalized 2D-band peak position; (b) normalized G-band peak position; (c) normalized FWHM of the 2D-band; (d) normalized FWHM of the G-band; (e) normalized intensity ratio ($I_{2D}/I_G$); (f) normalized intensity ratio ($I_D/I_G$).

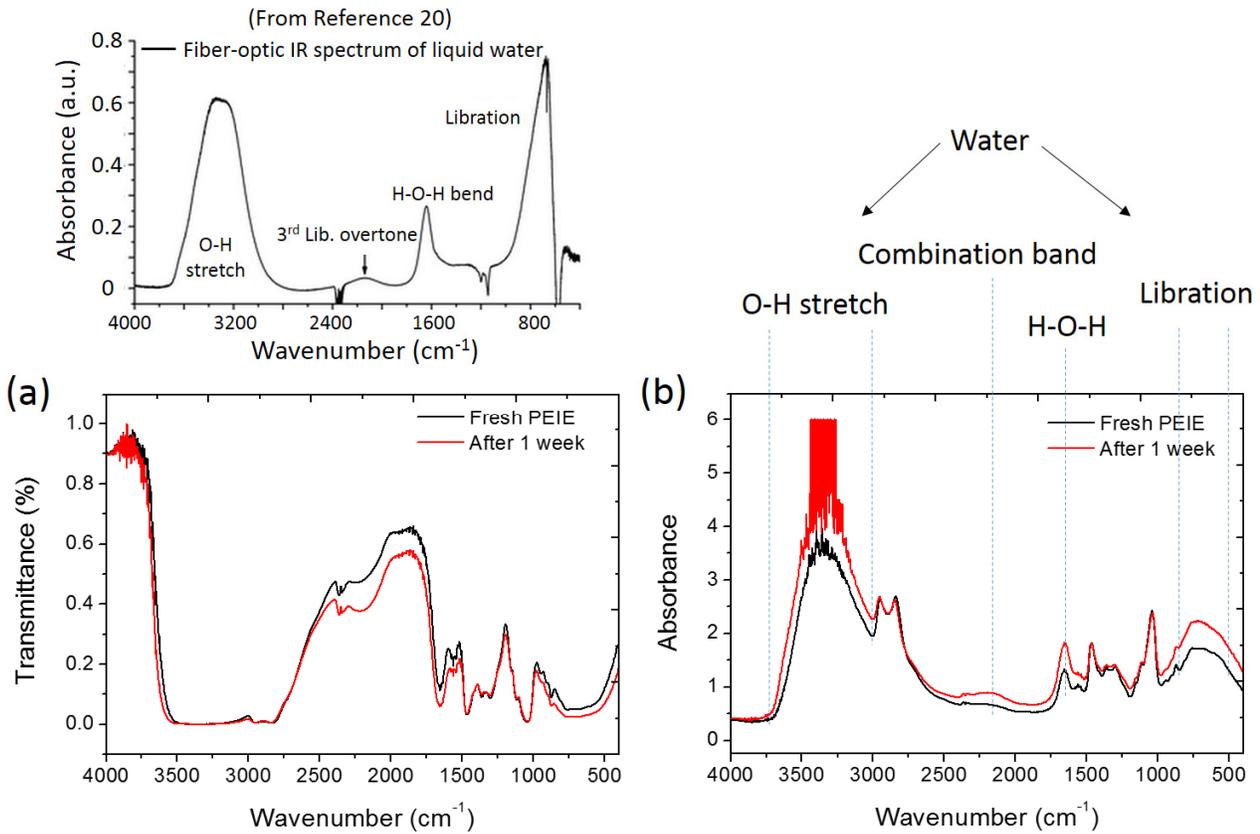

FIG. 7. Time evolution of the FT-IR spectra of PEIE: (a) Comparison of the transmission spectra taken on freshly prepared PEIE and on the same PEIE sample aged in air for one week. (b) Comparison of the absorbance spectra taken on freshly prepared PEIE and on the same PEIE sample aged in air for one week. Three specific molecular modes associated with water are indicated according to the spectrum of water (shown above (a)) taken from Reference [20].

In comparison with doped monolayer graphene in air, the aging effect on the XPS under 100% humidity appeared to be significantly worsened, as shown in Fig. 8(c) for the C-1s spectrum of the hole-doped monolayer graphene that revealed a binding energy increased up to 0.4 eV after 30 days. The S-2p spectrum also exhibited an increase in binding energy up to 0.4 eV after 30 days, indicating that the Fermi level of graphene has been downshifted by 0.4 eV. Similarly, for electron-doped monolayer graphene, the saturated moisture gave rise to 0.4 eV downshift in the binding energies of both the C-1s and N-1s core electrons (Fig. 8(d)), which doubled the energy shift of electron-doped monolayer graphene in air. Moreover, excess spectral features at lower binding energies of the C-1s spectrum were found to increase with time under 100% humidity, which suggested excess electron doping to graphene from saturated moisture.

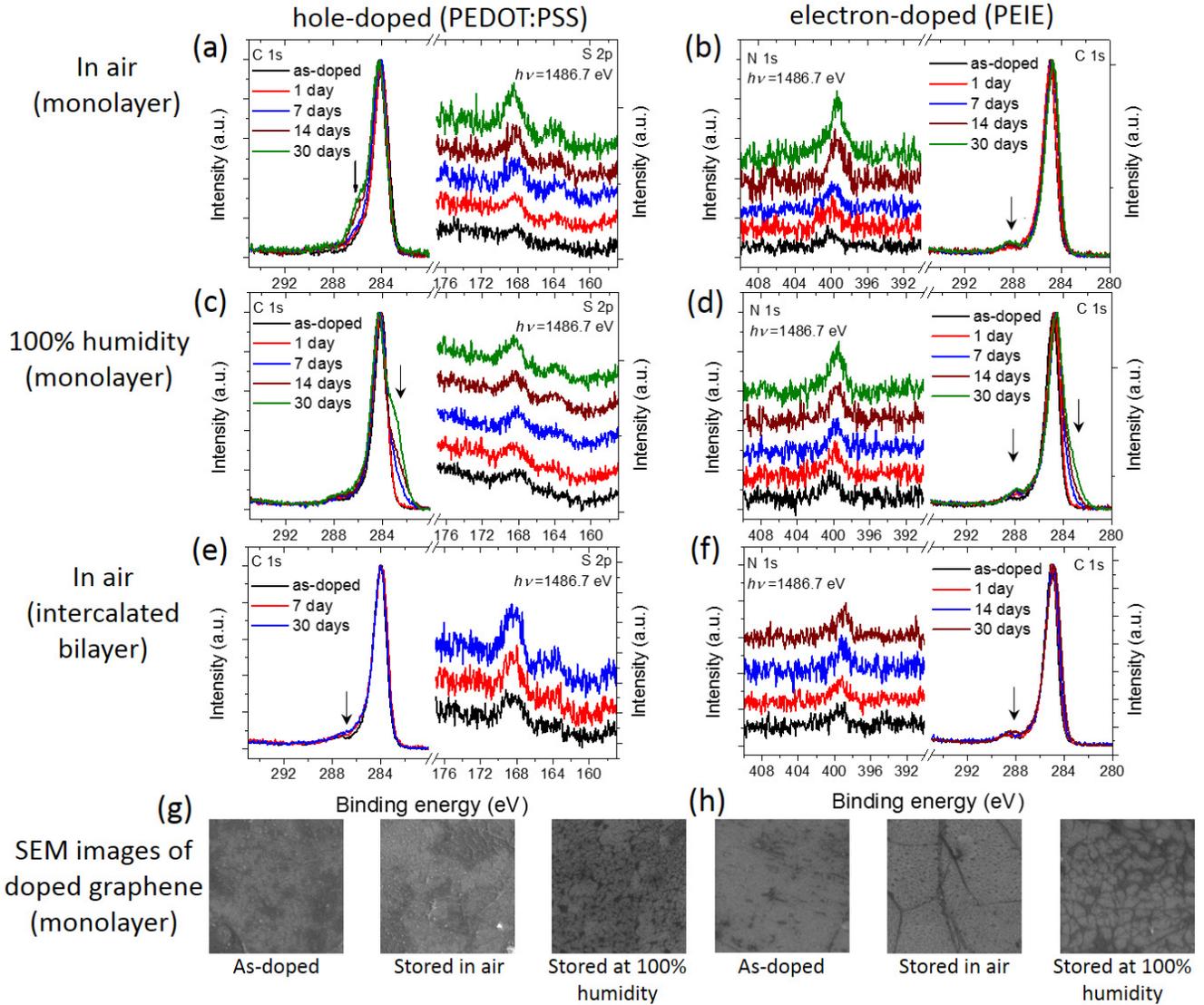

FIG. 8. Comparison of the XPS and SEM images of doped graphene under different conditions: (a) C-1s and S-2p core electron spectra of hole-doped monolayer graphene in air, showing enhanced intensities of the sp$^2$ hybridized C-OH (at 285.7 eV) and the sp$^3$ C-O (at 286.6 eV) components with time. (b) C-1s and N-1s core electron spectra of electron-doped monolayer graphene in air, showing enhanced intensity of the C-C=O component at 288.7 eV. (c) C-1s and S-2p core electron spectra of hole-doped monolayer graphene in 100% humidity. Excess spectral features below the C-1s peak were found to increase with time. (d) C-1s and N-1s core electron spectra of electron-doped monolayer graphene in 100% humidity. Excess spectral features below the C-1s peak were also found to increase with time. (e) C-1s and S-2p core electron spectra of PEDOT:PSS intercalated bilayer graphene in air. (f) C-1s and N-1s core electron spectra of PEIE intercalated bilayer graphene in air. (g) SEM images of hole-doped monolayer graphene over a (4 × 4) μm$^2$ area, from left to right: as-doped in air, after 30 days in air, and after 30 days in 100% humidity. (h) SEM images of electron-doped monolayer graphene over a (4 × 4) μm$^2$ area, from left to right: as-doped in air, after 30 days in air, and after 30 days in 100% humidity.

To better understand the physical origin of excess electron doping under 100% humidity, we performed simulations based on the density functional theory (DFT)[23,24] to investigate the electronic interaction between carbon atoms in graphene and $H_2O$ molecules. As detailed in the supporting information, we find that when excess $H_2O$ molecules are placed close to the vicinity of graphene, they prefer to forming a monolayer that stretches parallel to the graphene sheet (Appendix and Figure 9), and there is a net electron transfer to the carbon atoms on graphene from the $H_2O$ molecules, with the hydrogen atoms of the water molecules being closer to the carbon atoms in graphene than the oxygen atoms (Appendix and Figure 10). Since carbon atoms with $H_2O$ molecules attached would be surrounded by more negative charges as compared to normal carbon atoms in graphene without $H_2O$ molecules, the electron binding energy of $H_2O$-attached carbon atoms would be lower than that of the rest of carbon in graphene due to this physisorption process. As a result, there would be extra features appearing at the lower binding energy side of the XPS C-1s spectra, which is consistent with our experimental findings in Figs. 8(c) and 8(d). We further remark that this extra electron doping from saturated moisture to graphene is independent of the hygroscopic effects of the organic dopants that mediate chemical bonds to doped graphene, and is only significant under the 100% humidity condition.

In contrast, the aging effect on the XPS data of intercalated bilayer graphene appeared to be significantly mitigated relative to those of the monolayer graphene, as manifested in Fig. 8(e)-(f). For PEDOT:PSS intercalated bilayer graphene, both binding energies of the C-1s and S-2p core electrons appeared to be invariant up to 30 days, and the $sp^2$ hybridized C-OH and the $sp^3$ C-O components only exhibited slight increase relative to those of the hole-doped monolayer graphene (Fig. 8(a)). In the case of PEIE intercalated bilayer graphene, both the binding energies of the C-1s and N-1s core electrons were downshifted by 0.1 eV in air after 30 days, which was smaller than the downshift of 0.2 eV for monolayer graphene in air after 30 days, suggesting that the influence of aging effects on the work function was reduced. Moreover, the increase of the C-C=O component with time was also weakened slightly for the intercalated bilayer graphene. These findings of weakened aging effects on intercalated bilayer graphene relative to doped monolayer graphene are consistent with those revealed from the Raman spectroscopic studies.

While the spectroscopic studies described above revealed the averaged effects of different conditions on the aging of graphene samples, additional investigation of spatially resolved SEM images of these graphene samples as a function of time can provide more information about the underlying mechanism of the aging effect. As exemplified in Fig. 8(g)-(h), both of the as-doped hole- and electron-doped monolayer graphene revealed relatively clean SEM images other than a few line defects, whereas dark spots appeared to develop along the defect lines and more defect lines proliferated after 30 days, particularly for samples stored in 100% humidity. We believe that the increasing dark spots and defect lines were associated with the development of degraded graphene, although we could not independently verify the local Raman spectra of the small dark spots and narrow lines due to the limited spatial resolution (~ 2$\mu$m) of our Raman spectroscopy.

## IV. DISCUSSION

Our spectroscopic studies of the aging effect of doped graphene suggest that the primary cause for degradation is due to the reaction of the hygroscopic organic dopants with environmental moisture, which results in reduction of carrier densities in the doped graphene and also creates excess edge states due to increasing damages propagated from the grain boundaries. The apparent difference in the severity of the aging effect between the hole- and electron-doped graphene may be attributed to the different numbers of reaction sites in the organic dopants rather than possible asymmetric behavior between the valence and conduction bands of graphene.

On the other hand, the improved spectral stability of doped graphene in the intercalated bilayer configuration suggests that the enclosure of organic dopants between two monolayers graphene can protect the organic molecules from fast environmental degradation. This finding differs from previous reports that primarily focused on the protective effect of stacking multiple graphene layers on top of the surface of the material of interest.[6] We believe that the configuration of intercalated bilayers can provide additional environmental protection because of the graphene layer between the intercalant and the underlying substrate: Our studies of all doped graphene samples were transferred to $SiO_2$/Si substrates, which exhibited relatively rugged surfaces[25] and so could result in trapping of or pathways for environmental gases/liquids. Therefore, the insertion of a graphene monolayer at the interface of an organic dopant and the $SiO_2$/Si substrate could significantly reduce the aging effect and improve the long-term stability of organic optoelectronic devices.

## V. CONCLUSION

We have conducted systematic time-dependent Raman, XPS, UPS and SEM studies of organically doped graphene samples up to 30 days as a function of humidity, the number of graphene layers and the type of organic dopants. Detailed spectral analysis revealed that the aging effect of doped graphene was primarily associated with continuing reaction of organic dopants with environmental moister, which mediated excess chemical bonds to graphene and continuing reduction of doped carriers in graphene, and also induced proliferating damages along the grain boundaries of graphene. Additionally, high humidity can result in physisorption of a thin layer of water on the surface of graphene, giving rise to reduction in the electron binding energy of graphene. On the other hand, organic dopants in the intercalated bilayer configuration become better protected against environmental degradation. Therefore, low humidity, low-defect large-area graphene sheets and the addition of an extra graphene layer at the interface between the organic dopant and the underlying substrate can provide significant long-term stability for organically doped graphene, lending better reliability for low-cost and flexible graphene-based electronic and optoelectronic devices.

**ACKNOWLEDGMENTS**

The work at Caltech was jointly supported by the National Science Foundation through the Institute of Quantum Information and Matter (IQIM) at Caltech, the Sobhani Foundation, and the Moore Foundation. JKC and CIW acknowledge the support of Dragon Gate Program by the National Science Council in Taiwan. We thank Professor George Rossman for the use of Raman spectrometer, and acknowledge the use of XPS facilities at the Beckman Institute at Caltech.

**APPENDIX: DENSITY FUNCTIONAL THEORY SIMULATIONS OF THE EFFECT OF WATER MOLECULES ON GRAPHENE**

We investigate how the presence of water molecules may influence the electronic structures of the carbon atoms in a graphene sheet by using a semi-empirical hybrid method known as the B3LYP method,[26] which was derived from the density functional theory (DFT).[23,24] The B3LYP method can provide high accuracy for molecular computation, and all optimization by the B3LYP method in this work was made by using a quantum-chemical calculation package, Gaussian 03 (Gaussian, Inc.), with a 6-31G(d,p) basis set.

For meaningful results attainable with a reasonable computation time, we use the ``coronene'' (see Figure 9) molecule to simulate reduced graphene. The optimized electronic structure of the system with the Mulliken charge[27] of each carbon atom specified is shown in Figure 9. Next, we add $H_2O$ molecules to the system and minimize the Gibbs free energy. We find that the Gibbs free energy is negative in the presence of $H_2O$ molecules and decreases with the increasing number of $H_2O$ molecules, which implies that it is energetically favorable for $H_2O$ molecules to react with graphene. The resulting Mulliken charge distributions for individual atoms in the assembly of reduced-graphene and 5 $H_2O$ molecules are shown in Figure 10(a). We note that the $H_2O$ molecules prefer to forming a monolayer stretching parallel to the reduced graphene rather than stacking above other $H_2O$ molecules, and the hydrogen molecules are closer to the graphene than the oxygen molecules, as exemplified in Figure 10(a) and 10(b) for the top view and side view, respectively, for a system of five $H_2O$ molecules and a coronene. The total Mulliken charge of $H_2O$ is negative, suggesting that $H_2O$ molecules lose electrons to carbon atoms, leading to lower binding energies for the C-1s spectrum, which is consistent with our XPS data.

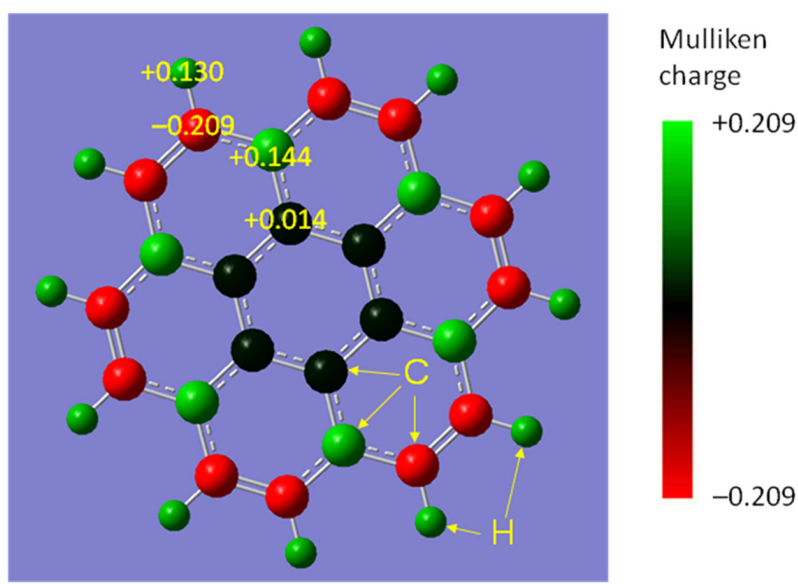

FIG. 9. The Mulliken charges of carbon atoms in the "coronene", a piece of reduced graphene terminated by hydrogen atoms, from DFT calculations: The Mulliken charge for carbon atoms is represented in color scales. We note that the Mulliken charge for carbon atoms at equivalent positions is identical by symmetry.

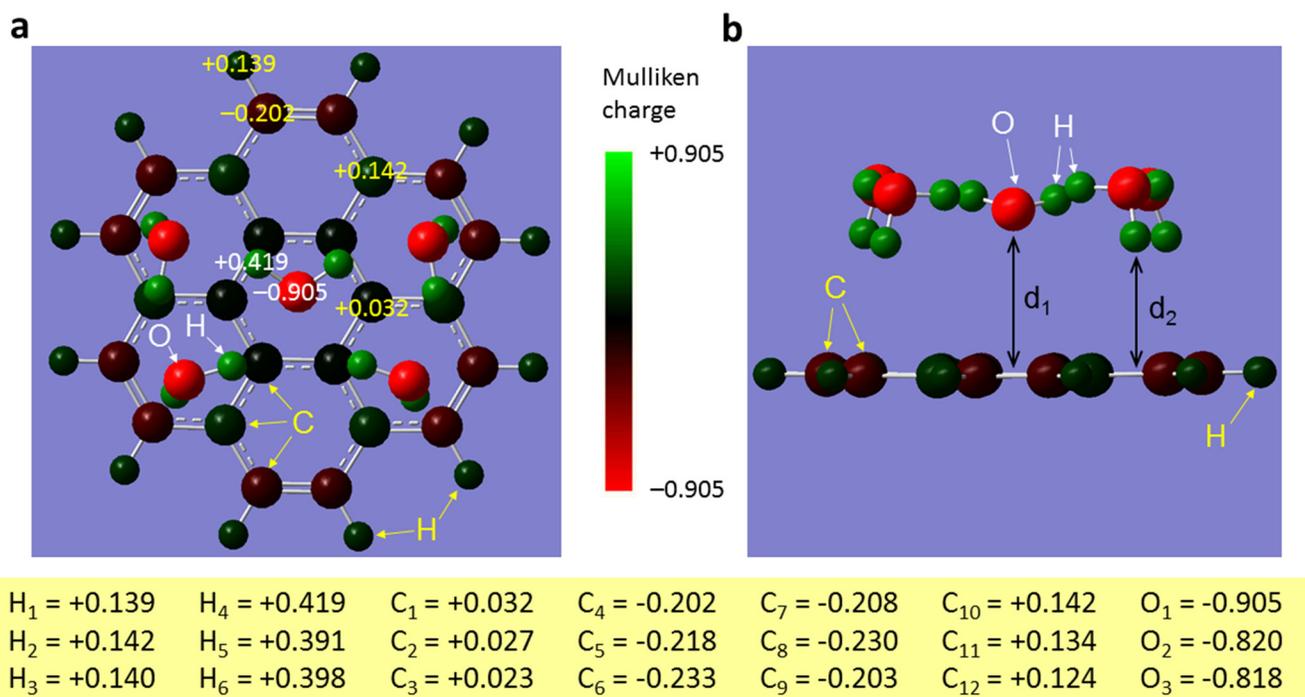

FIG. 10. The preferred atomic arrangement of 5 $H_2O$ molecules near the reduced graphene, from DFT calculations: (a) Top view of the atomic arrangements and the Mulliken charge distributions for individual atoms, where the specific Mulliken charges for inequivalent 12 carbon atoms (denoted as $C_i$ where $i = 1 -- 12$), 6 inequivalent hydrogen atoms (denoted as $H_j$ where $j = 1 -- 6$) and 3 inequivalent oxygen atoms (denoted as $O_k$ where $k = 1 -- 3$) are listed. (b) Side view of the atomic arrangements, showing the $H_2O$ molecules forming a monolayer above the reduced graphene, with hydrogen atoms closer to the graphene sheet than oxygen atoms and a net transfer of electrons from the $H_2O$ molecules to the graphene sheet.